\documentclass[rnaas]{aastex63}
\usepackage{amsmath}
\usepackage{multirow}

\def \fermi{{\it Fermi}-LAT~}
\graphicspath{{./}{figures/}}

\begin{document}

\title{On the gamma-ray emission of W44 and its surroundings}

\correspondingauthor{Giada Peron}
\email{giada.peron@mpi-hd.mpg.de}

\author{Giada Peron}
\affiliation{Max-Planck-Institut f{\"u}r Kernphysik, P.O. Box 103980, 69029 Heidelberg, Germany}
\author{Felix Aharonian}
\affiliation{Dublin Institute for Advanced Studies, 31 Fitzwilliam Place, Dublin 2, Ireland}
\affiliation{Max-Planck-Institut f{\"u}r Kernphysik, P.O. Box 103980, 69029 Heidelberg, Germany}
\author{Sabrina Casanova}
\affiliation{Institute of Nuclear Physics, Radzikowskiego 152, 31-342 Krakow, Poland}
\affiliation{Max-Planck-Institut f{\"u}r Kernphysik, P.O. Box 103980, 69029 Heidelberg, Germany}
\author{Roberta Zanin}
\affiliation{CTA observatory, Via Piero Gobetti 93/3
40129 Bologna, Italy}
\affiliation{Max-Planck-Institut f{\"u}r Kernphysik, P.O. Box 103980, 69029 Heidelberg, Germany}
\author{Carlo Romoli}
\affiliation{Max-Planck-Institut f{\"u}r Kernphysik, P.O. Box 103980, 69029 Heidelberg, Germany}

\begin{abstract}
We present the analysis of 9.7 years \fermi{} data of the middle-aged supernova remnant W44 and the massive molecular gas complex that surrounds it.  We derived a high-quality spectral energy distribution of gamma-radiation of the shell over three decades. The very hard spectrum below 1~GeV supports the earlier claims regarding the hadronic origin of radiation.  We also confirm the presence of two extended  $\gamma$-ray structures located at two opposite edges of the remnant along its major axis. {  Based on the high-resolution gas maps, we demonstrate that  the gamma-ray structures are caused by the enhanced cosmic-ray density rather than the gradient of the gas distribution}. We argue that the {  revealed cosmic-ray ``clouds'' suggest an anisotropic character of the escape} of high-energy particles from the shell along the magnetic field of the remnant.  
\end{abstract}

\section{Introduction} 
In the current paradigm of Galactic cosmic rays (CRs), supernova remnants (SNRs) hold a central role. For decades, they have been believed to be CR factories and major contributors to the fluxes of locally observed CRs.  It has been recognized long ago that the detection of gamma-rays from SNRs - either isolated \citep{DAV} or interacting with massive molecular clouds \citep{ADV} - could provide key insights into nonthermal processes in these objects. Indeed, the recent reports on the detection of many SNRs in high and very high energy (VHE) gamma-rays unequivocally demonstrate the effective acceleration of CRs in SNRs.  At the same time, these observations pose certain concerns regarding the adequacy of interpretation of multiwavelength data within the standard schemes of particle acceleration and radiation. 

 The impressive list of reported gamma-ray-emitting SNRs includes several famous representatives of both young and middle-aged remnants. This is quite essential for understanding the dynamics of two competing processes of particle acceleration and escape in SNR. In this regard, the middle-aged SNR W44, reported as a strong high energy (MeV/GeV) gamma-ray emitter \citep{agileW44,w44fermi}, is of special interest.  GeV $\gamma$-ray observations of W44 with AGILE and \fermi \citep{agileW44,w44fermi}  revealed an evidence for proton acceleration in the remnant.  Moreover. this source has the \textquotedblleft right" age,  $\sim$10,000 yr  \citep{snrcat},  for investigating the escape of accelerated particles. It is expected, in fact, that a significant fraction of accelerated particles have already left the remnant and entered the surrounding dense environment.  The gas complex, in which W44 resides, is massive enough that \fermi should be able to detect the diffuse $\gamma$-ray emission arising 
 from interactions of escaped relativistic particles. 
 Indeed, \cite{uchiyamaW44} have unveiled an extended $\gamma$-ray emission surrounding the shell of W44 and interpreted it as the escape of CRs from the shell. Combined with observations of $\gamma$-rays directly from the remnant, this diffuse component of radiation contains unique information about the history of particle acceleration and escape,  provides an unbiased estimate of the total energy released in the form of accelerated particles, and allows us to reconstruct the initial (acceleration) spectrum.   

In this Letter, we analyze the $\gamma$-ray emission in the region of W44,  based on almost 10 years of Fermi-LAT Pass8 data.  The significantly enhanced photon statistics allowed us to conduct new detailed spectral and morphological studies of both W44 itself and the regions beyond the remnant.  We briefly discuss the implications of these results concerning the origin of the gamma-ray emission from W44 and the character of the escape of accelerated particles from the remnant. 

\section{$\gamma$-Ray observations}
We analyzed 9.7 years of \fermi{} Pass8 data in a 10$^\circ$ region of interest (ROI), centered on W44 (l=34.60$^\circ$, b=-0.36$^\circ$). We selected ’FRONT+BACK’ events (evtype=3), with zenith angles larger than 90$^\circ$, in order to avoid the Earth Limb, and imposed DATA\_QUAL==1 \& \& LAT\_CONFIG==1. We used the latest released models for the galactic (\texttt{gll\_iem\_v07.fits})  and extragalactic background  (\texttt{iso\_P8R3\_SOURCE\_V2\_v1.txt}), and  included in the model the gamma-ray sources from the 4th \fermi{} catalog 4FGL \citep{4fgl}.

We remodeled the morphology and the spectral shape of  the closest 4FGL sources ($\lesssim$ 1$^\circ$) to the remnant. The two newly catalogued extended sources, 4FGL J1857.7+0246e and 4FGL J1852.4+0037e, that coincide with the pulsar wind nebula candidate HESS J1857+026 and the region of the SNR Kes79, are of special interest. Further details are reported in the Appendix. 

Finally, we modeled the spatial and energy distributions of gamma-ray emission for both  W44 and the surrounding regions, as explained in the following sections. 

\subsection{The Supernova Remnant}
W44 is a  {\it mixed-morphology} SNR \citep{mixed}. At radio wavelengths, W44 shows an elliptical shell-like structure. Inside the remnant,  a pulsar, PSR B1853+01, and its nebula have been detected in radio and in X-rays. Both do not show high-energy counterparts \citep{w442010}. We performed a new investigation of the morphology of the remnant, applying  the recent \fermi background model. For morphological studies, only gamma-rays with energy exceeding 1 GeV have been analysed, because at these energies the point-spread function (PSF) is sufficiently small ($\sigma_{PSF} \leq 0.3^\circ$).   Previous studies \citep{w442010} reported that in $\gamma$-rays W44 shows an elliptical ring shape.  We confirm the same morphology,  namely an elliptical ring of size ([0.18$^\circ$, 0.3$^\circ$] and [0.13$^\circ$, 0.22$^\circ$]).  The gamma-ray ring closely matches the image of the shell obtained in radio, in particular  by the NRAO VLA Sky Survey at 1.3628-1.4472 GHz  frequencies \citep{nvss}  and the 
THOR survey at 1.4-1.8 GHz \citep{thor}.

After the morphology was fixed, we derived the spectral energy distribution (SED) of gamma-rays in the whole energy range, starting from $\sim$60 MeV, and then explored the origin of this $\gamma$-ray emission with the \texttt{naima} package \citep{naima}. 

\begin{figure*}[ht!]
    \centering
    \includegraphics[width= 0.7\linewidth]{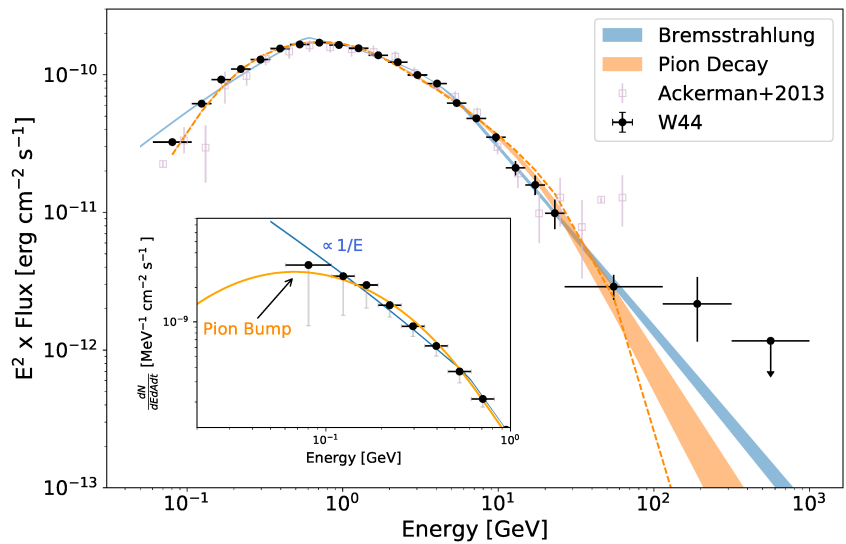}
    \caption{The Spectral energy distribution of $\gamma$-rays from the Supernova Remnant W44 obtained with 9.7 years Pass8 data. The SED from a previous analysis by \cite{w44fermi} is indicated as light purple squares. We over-plot the gamma-ray emission modeled with naima for a hadronic and a leptonic scenario.  The {  orange} zone shows the 1-sigma confidence level for the derived gamma-ray spectrum  produced in interactions of accelerated protons and nuclei with the surrounding gas. The derived proton distribution is a broken power-law in total energy with the indices $\alpha_1= 2.40 \pm 0.02$ and  $\alpha_2= 3.87 \pm 0.14$  below and above $E_b=39\pm 3$~GeV. {  A power-law with exponential cutoff at 71$\pm 6$ GeV and $\alpha=2.30\pm 0.02$ can also interpolate the data as the orange dashed-line shows}.  The {  blue} zone represents the confidence levels of the gamma-ray spectrum  obtained  by the  bremsstrahlung fit assuming a broken power-law distribution of electrons with indices $\alpha_1=2.32 \pm 0.03$  and $\alpha_2= 3.39\pm 0.07 $ and a break at  $E_b=6.1 \pm 0.6$~GeV, and a low energy cutoff at $E^e_{min}=600 $~MeV.  The inset  panel shows the {  differential spectrum below 1 GeV.{Here the hadronic model has been extended down to 20 MeV to show the characteristic pion bump. {   Light grey error-bars indicate the systematic uncertainties (see the text). Statistical error bars instead are not visible on this scale.} } } }
    \label{fig:w44SED}
\end{figure*}

The adequate photon statistics allows us to derive high-quality SEDs over three decades, from $\sim$100 MeV to $\sim$100 GeV,  as shown in Fig.\ref{fig:w44SED}. Generally, our results agree with the spectra reported in previous studies; however we extended the spectrum, by adding statistically significant points both at the lowest and at the highest edges of the energy range. 
We evaluated the systematic uncertainties due to the diffuse background model, by comparing the result obtained with the latest version of the galactic background, and the former one (\texttt{gll\_iem\_v06.fits}). {  The SED obtained with the latter is systematically lower.} The difference is higher  at the lowest energies where it is of the order of $\sim$60\%, whereas in the rest of the spectrum is $\lesssim 10 \%$ . 

{  The overall spectrum of W44 is presented in Fig.\ref{fig:w44SED} and can be interpolated by a log-parabola function: $$\frac{dN}{dE} = N_0 \bigg( \frac{E}{E_b} \bigg)^{-(\alpha +\beta \log(E/E_b))} $$ with the best-fit values $N_0= (9.5 \pm 0.1 ) \times 10^{-12}$ (MeV cm$^2$ s)$^{-1}$, $\alpha=2.57 \pm 0.01 $, $\beta = 0.235 \pm 0.005$  and $E_b=2.67  $ GeV. 
As one can notice, the differential spectrum, below a few hundred MeV is flat, ${\rm d} N/{\rm d} E \propto E^{-0.5}$ and correspondingly, the SED is very hard $\propto E^{1.5}$.  Such a sharp raise of the SED is naturally explained by the {\it pion bump} feature at $m_{\pi}/2 \approx 67$.5 MeV caused by the kinematics of the $\pi^0$-decay, as shown in the inset panel of Fig. \ref{fig:w44SED}.
\cite{w44agilefermi} have argued that a single leptonic scenario is incompatible with the radio synchrotron data. This is a viable,  although, strictly speaking, model-dependent approach, because it does not take into account that electrons producing gamma-rays could have a different origin than the electrons responsible for the radio emission. Conversely, the leptonic origin of gamma-rays could be unambiguously discarded by observations of low-energy gamma-rays with a differential spectrum harder than $E^{-1}$. The latter is the hardest possible bremsstrahlung spectrum that could be formed in the case of low-energy cutoff in the spectrum of electrons \citep{w44fermi,hb21}. Indeed, in Fig.\ref{fig:w44SED} we show that while the electron bremsstrahlung  can explain the high-energy spectral points quite well, at energies below a few hundred MeV it fails to fit the flattening of the flux, even if we assume a sharp low-energy (600 MeV) cutoff in the electron spectrum. Meanwhile, at such low energies, the second leptonic channel - the inverse Compton scattering -  is too inefficient.  All of these arguments favour the hadronic scenario. Unfortunately, the large uncertainties of the low-energy points prevent us from the robust rejection of the leptonic origin of gamma-rays. {  Nevertheless the systematic uncertainty points toward the hadronic interpretation of the emission}}

{ At energies above 10~GeV, the spectrum of W44 drops quickly, with a power-law  index of  $3.3 \pm 0.4$ 
(see Fig. \ref{fig:w44SED}). {  This could happen in} the case of inefficient diffusive shock acceleration with a Mach number $\ll 10$, as discussed in depth by \cite{w44reacc}. They proposed that the steep spectrum is a result of the cutoff due to the maximum energy of accelerated particles at $\leq 10$~GeV.  However, the introduction of such low cutoff contradicts the extension of the derived parent proton spectrum to energies up to 1 ~TeV, unless one assumes an effective reacceleration inside the shell. This would result in a hardening of the $\gamma$-ray spectrum at higher energies, as claimed above 30 GeV in \cite{w44reacc}. However, our study based on significantly larger photon statistics does not support the tendency of hardening of the $\gamma$-ray spectrum, at least until 100 GeV.  For this reason, we interpret the steep gamma-ray spectrum as due to energy-dependent escape of protons, implying that the highest-energy particles have already left the remnant.} 

We fit the $\gamma$-ray spectrum in the whole energy band from 100 MeV to 100 GeV with a pion decay model.  In this case, the low-energy part of the spectrum is perfectly explained by the interaction cross-section  close to the kinematic threshold. To explain the steep  spectrum at high energies, we need to introduce a break or a cutoff in the parent proton spectrum {  around few tens of~GeV. This can be realized by a {\it power-law spectrum with an exponential cutoff} ($\alpha= 2.30 \pm 0.02 $, $E_{co}= 71\pm 6$ GeV)  or by a {\it broken power law spectrum}  ($\alpha_1= 2.40 \pm 0.02$, $\alpha_2=3.87 \pm 0.14$, $E_b= 39\pm 3$ GeV ) (see Appendix). 

In Fig. \ref{fig:w44SED} is also shown the best bremsstrahlung fit calculated for a  broken power-law electron spectrum ($\alpha_1=2.31 \pm 0.03$, $\alpha_2=3.38 \pm 0.06$, $E_{\rm b}=6.1 \pm 0.5 \ \rm GeV$), with an additional assumption of sharp low-energy cutoff at $E^e_{\rm min}$=600~MeV.  }

In the energy range of protons 1--1000~GeV corresponding to the energy interval of detected $\gamma$-rays,  the total energy of CR protons  inside the remnant is estimated as  $W_p=  1.2 \times 10^{50} ({n_p}/{1 \  \mathrm{cm}^{-3}})^{-1} ({d}/{2.2 \  \mathrm{kpc}})^{2}$~erg.  In the case of electron bremsstrahlung the total energy of electrons needed to support the observed $\gamma$-ray flux is $W_e=1.1 \times 10^{49} ({n}/{1 \  \mathrm{cm}^{-3}})^{-1} ({d}/{2.2 \  \mathrm{kpc}})^{2} $ erg.  Assuming a gas density in the remnant, similar to what is measured in the surrounding $n \approx  10$~ cm$^{-3}$ , gives a rather reasonable estimate of cosmic-ray protons  still confined in the remnant, $W_p=  1.2 \times 10^{49}$~erg.

\subsection{The Surroundings  }
\begin{figure*}[ht!]
\centering
\includegraphics[width=1\linewidth]{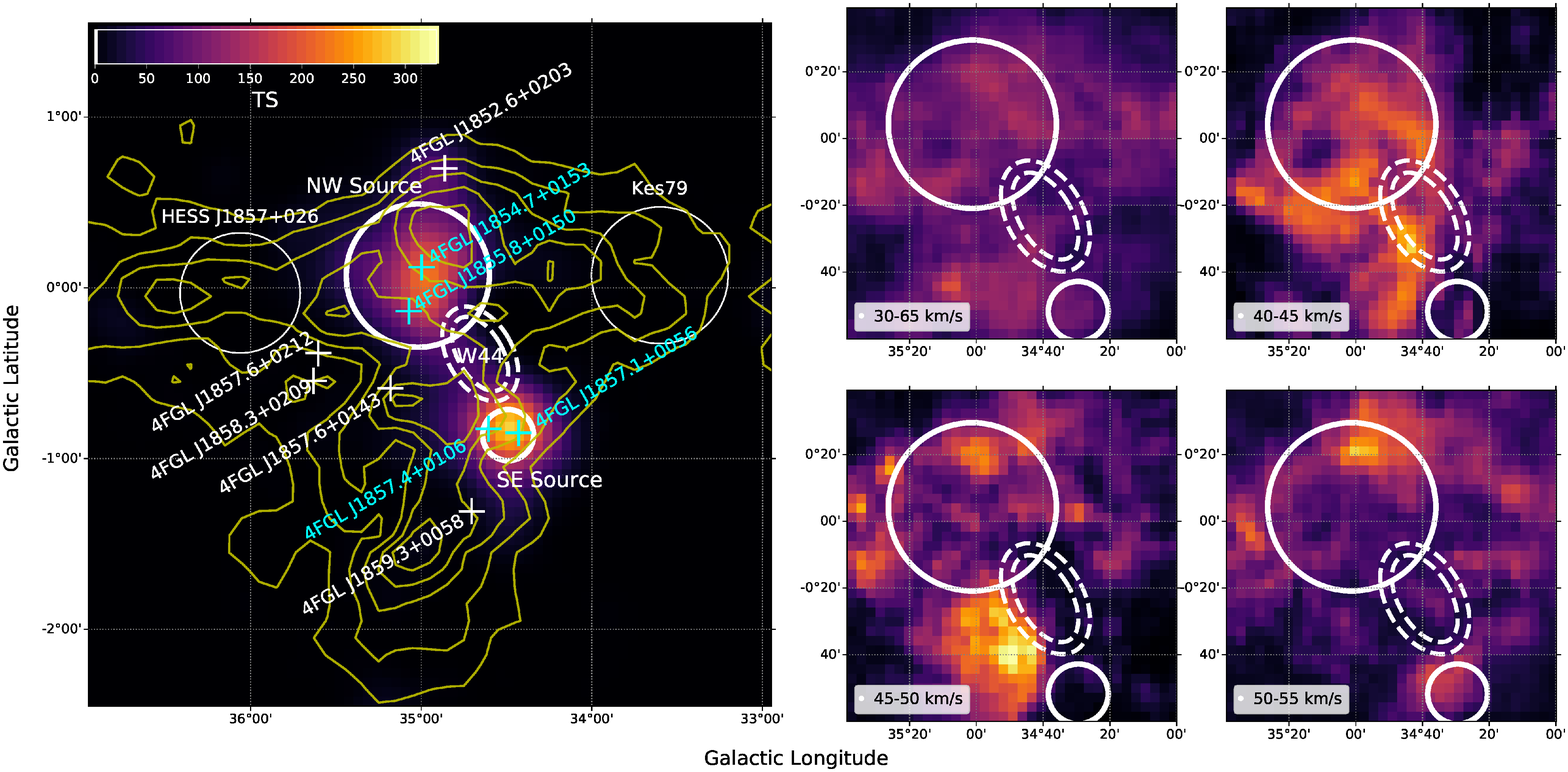}

\caption{On the left panel: Test Statistic maps of the two clouds. The position of all gamma-ray sources within 2 degrees from the remnant are shown. The white regions correspond to the final configuration of the sources. Cyan regions indicate the sources that have been removed from the model. The CO  contours from the 30-65 km/s interval  are also plotted in yellow \citep{Dame2001}.  On the right: the gas distribution from the Nobeyama telescope data \citep{fugin} in 4 different velocity ranges, as indicated in the labels. The white regions indicate the position of the  SE and NW $\gamma$-ray sources (solid), and of W44 (dashed). The gas density distribution is normalized to the mean value corresponding to each velocity range.}
\label{fig:clouds_ts}
\end{figure*}

To search for a possible diffuse $\gamma$-ray emission caused by interactions of accelerated particles,  escaped from the remnant, we analyzed the $\gamma$-ray emission originating in the surroundings of W44.  We firstly investigated the morphology of the emission. For that, we restricted the analysis to gamma-rays of energies higher than 1~GeV, to minimize the source confusion and to take advantage of the better PSF. We removed all 4FGL sources within 2-degrees from the center of the SNR  and re-modelled them. To evaluate the best configuration, we used the Aikake iformation citerion \citep{aic}.  We found two extended $\gamma$-ray sources, that overlay with the sources reported earlier by \cite{uchiyamaW44}, but with a slightly different extension. We call them {\it SE-Source} and {\it NW-Source},  to indicate that they are located at the south-east and north-west edges of the SNR (see
 Fig. \ref{fig:clouds_ts}).  In the 4FGL catalog, at the position of the clouds, four unidentified point-sources are included, namely J1857.4.0126, J1857.1+0056 in the South and J1855.8+ 0150 and J1854.7+0153 in the North. The AIC method favours, for both regions, a one-disk  configuration.  The fitted angular extensions of the $\gamma$-ray structures are $(0.15\pm 0.02) ^\circ$ and $(0.42\pm 0.03) ^\circ$ for the SE- and the  NW-  Source, respectively.

In order to find possible associations with gas complexes, we investigated the gas distribution {  around} the remnant, which has been located at v=45 km/s, thanks to OH maser observations \citep{maser}. We studied the gas distribution within 2 degrees from the remnant, in the velocity range  [30-65] km s$^{-1}$, as shown in Fig. \ref{fig:clouds_ts}. Generally,  the gas distribution in the the CO-map in this region looks quite uniform. However, the analysis of the high resolution  $^{12}$CO data obtained with the Nobeyama telescope  allowed \cite{seta98} to divide this gas complex in six smaller molecular clouds with masses from 0.3 $\times 10^5$ M$_\odot$ to 3 $\times 10^5$ M$_\odot$.  The newly identified extended $\gamma$-ray sources {  show no direct correspondence between the gas distribution and the location of gamma-ray emission, although some of these molecular clouds partially overlap with them}. SE-Source appears very close to GMC G34.6-0.7(V=53), which, according to \cite{seta98}, shows a hint of interaction with the remnant. The NW-Source partially overlays with another gas cloud,  GMC G35.0+0.3(V=50).  Otherwise, the gas density in the region of NW-Source is close to the average density of the region surrounding W44.  It is also interesting to notice, that we see no $\gamma$-ray signals from other closer and/or more massive clouds identified by \cite{seta98}, {  like GMC G34.8-0.6 (V=48 km/s), that is closer to the shell and has a mass  2--10 times larger than the gas inside the SE- and NW- sources. As shown in Fig.\ref{fig:clouds_ts}, the gas density in the region of the SE-  and the NW- Source is similar to the average value,
In both regions,
$n_{SE}\sim n_{NW} \sim 10$ cm$^{-3}$. We derived the mass of the gas, within the regions defined by the gamma-ray emission and in the velocity range of the SNR ([30,65] km/s) from the CO-map of \cite{Dame2001}}. They amount to   $M_{SE}= 0.4 \times 10^5 M_{\odot} \ (d/2.2 \  \mathrm{kpc})^2 $  and   $M_{NW}= 2 \times 10^5 M_{\odot} \ (d/2.2 \  \mathrm{kpc})^2 $ respectively.

\begin{figure*}

\includegraphics[width=1.\linewidth]{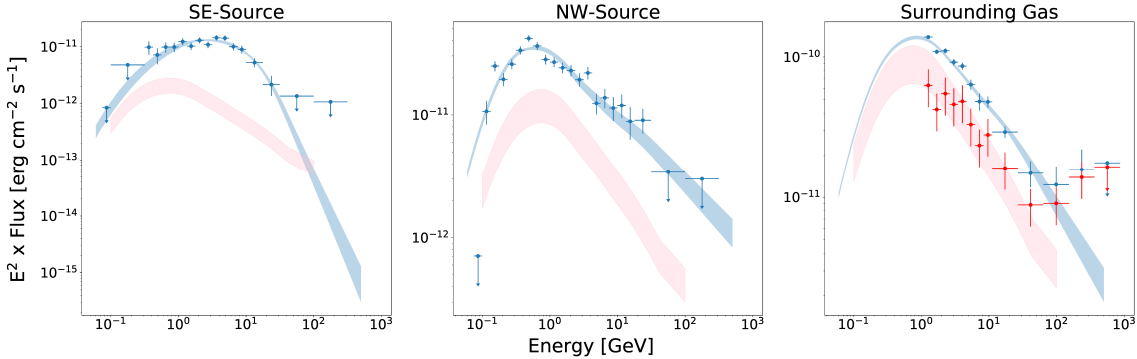} 
\caption{Spectral Energy Distributions of the SE, NW sources, and the entire gas complex (blue points). 
The blue zones are the best fits obtained with the {\it naima} code under the assumption of hadronic origin of $\gamma$-rays. The red zones are the $\gamma$-ray fluxes expected from the sea of CRs interacting with the gas complex.
For the gas complex we show the spectral points after subtraction of the fluxes of $\gamma$-rays linked to
the CR sea (red points)}
\label{fig:sed_surr}
\end{figure*}

{  In Fig. \ref{fig:sed_surr}, we compare the $\gamma$-ray fluxes of the SE and NW Sources with the fluxes of $\gamma$-rays induced by the  CR \textquotedblleft sea" , the large-scale ($\geq 10$~kpc) 
cosmic-ray population contributed by all galactic CR factories. The CR "sea" 
permeates the entire Galactic Disk with an average density close to the  local CR flux. Here we assume that the flux of the CR sea coincides with the fluxes reported by the AMS-02 experiment \citep{ams}.  {  In the computations, we assumed a nuclear enhancement  factor of the gamma-ray flux of 1.8 \citep{Mori2009,pp_parametrization} for CRs below 100 GeV. The latter takes into account the contribution of nuclei heavier than the hydrogen both in  CRs and in the interstellar medium (ISM).}}

As one can see,  the fluxes of the SE- and NW- Source are higher by a factor $\sim$ 8 and $\sim $ 3 compared to the CR sea.   This difference is unlikely to be due to the non-traced ({\it a.k.a. dark}) gas, that cannot exceed  50\%  \citep{codark} of the total mass.  Thus, we infer that there must be a concentration of high-energy particles in these regions. The spectral shape of these sources also differs from the spectrum of the cosmic ray sea.  The NW-Source spectrum reaches its maximum around $\sim$500 MeV,  like the CR-sea,  but is significantly harder at high-energies. The SE-Source spectrum is steeper than the cosmic ray sea at high energies but its SED peak is shifted towards a few GeV.  This implies a noticeable suppression of both from low- and high-energy protons. 

\begin{figure}
    \centering
    \includegraphics[width=0.5 \linewidth]{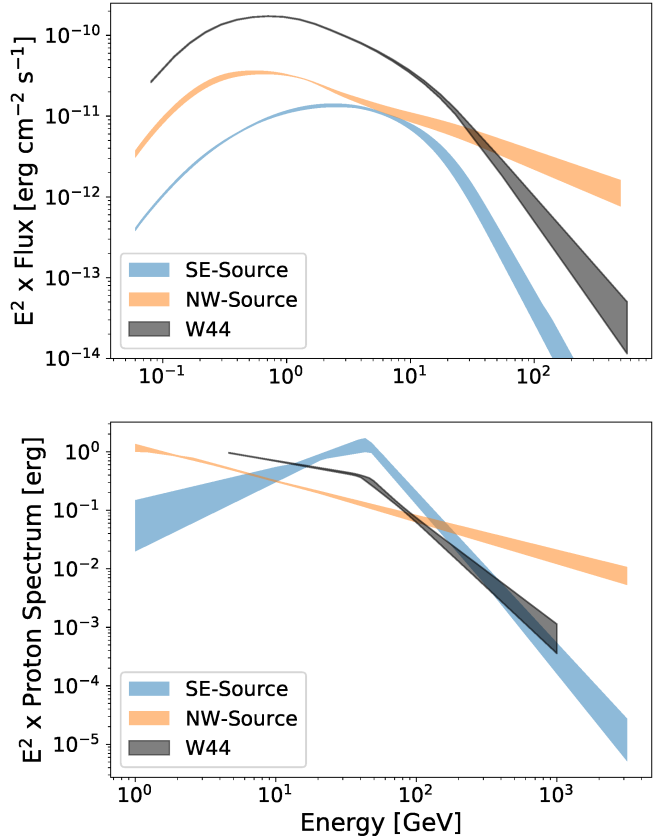}
    \caption{Upper panel: comparison of the SED of W44 with the SED of the $\gamma$-ray clouds. Lower panel: comparison of the derived proton distribution of the above mentioned objects.}
    \label{fig:spec_compare}
\end{figure}

 We have calculated the proton spectra inside  NW- and SE-Source using the code \texttt{naima} \citep{naima}, under the assumption  that the main emission mechanism is the decay of neutral pions, produced in interactions of accelerated protons and nuclei with the surrounding gas. The parameters characterizing these spectra are shown in Table \ref{tab:cr_values}.  In Fig. \ref{fig:spec_compare} we compare the SEDs and the proton distributions of the remnant to the ones of the two extended sources. At low energies, the spectrum of the SE-Source is harder than the spectrum of W44. This might be an indication that the low-energy particles produced by W44 have not yet reached the location of the gamma-ray source. On the other hand,  the very  soft spectrum of protons above few tens of GeV, suggests that most of the very high energy particles already left the location of SE-Source.  
 
 {  Although the distances from the centers 
 of two sources from the remnant are comparable ($\sim 25$ pc), the spectrum of NW-Source  is different from the SE-Source. At low energies, it is rather similar to the spectrum of $\gamma$-rays from the remnant.  In principle, this could be caused by
 the contamination introduced by W44 itself, given that NW-Source extends up to the edge of  the remnant. At higher energies, because of the better  angular resolution of \fermi,  an effect of contamination can be excluded. Indeed,
 one can clearly see that the spectrum of the NW source is significantly harder that the spectrum of the remnant.} 
 
Finally, we investigated the diffuse gamma-ray emission originating from the molecular complex around the Supernova Remnant. W44 is surrounded by a huge molecular gas complex  \citep{Dame2001}; within a 2$^{\circ}$ area the total mass is estimated to be $M_{surr} \approx 1.5 \times 10^6$ M$_\odot$. The corresponding  $M_5/d_{\rm kpc}^2$ parameter,  which characterizes the $\gamma$-ray fluxes of molecular clouds embedded in the CR sea, at the distance of the remnant ($d=2.2$ kpc), is about 3. This exceeds the threshold of detection of $\gamma$-rays from \textquotedblleft passive" GMCs by {\it Fermi}-LAT \citep{sea18}. Any deviation from the minimum level of $\gamma$-ray flux set by the CR sea, would imply presence of an additional CR component consisting of particles that already escaped  W44. To evaluate the $\gamma$-ray flux from this complex we constructed a {\it customized} model for the background. We made the spatial templates using the radio data-cubes of the interstellar medium (CO and HI) \citep{Dame2001,HI4PI} and a {\it GALPROP} based model for the inverse Compton component \citep{galprop}. 
Generally, in any background template, the $\pi^0$-decay,  the electron bremsstrahlung, and the inverse Compton diffuse gamma-ray components should be included. At energies below 1 GeV,  the contributions of these components are comparable. However, at lower energies the uncertainties related to electrons and protons are large which does not allow us to construct a reliable template. For this reason, we use the {\it customized} background model only for energies above 1 GeV.  At these energies, the contribution of the electron bremsstrahlung becomes negligible compared to the $\pi^0$-decay and the inverse Compton. The advantage of the {\it customized} background model is that it allows us to extract and analyze the cube of gas in the velocity range of the remnant. This is not possible in the case of the standard \fermi{} galactic background since it is constructed as a two-dimensional template and therefore the components along the line of sight cannot be separated. 

We excluded from the background gas in a box of $\Delta l \times \Delta b\times \Delta v=4^\circ \times 4^\circ \times 30$ km s$^{-1}$, centered on W44 and analyzed this region as a separate source. We didn't consider the atomic gas as source, but only as background, as it is small in that region  compared to the molecular component (M$_{HI}< 30\%M_{H_2} $).

Figure \ref{fig:sed_surr} shows the measured $\gamma$-ray spectrum, together with the predicted one calculated for the mass of $M_{\rm surr}$ and the flux of  the nominal CR sea. One can see that, unlike the SE and NW sources,  the shapes of the derived and predicted $\gamma$-ray spectra of the entire gas complex are similar. On the other hand, the derived absolute flux exceeds the flux calculated for the CR sea by a factor of 1.5.  The deviation is not significant, it is comparable to the uncertainties in the mass estimates of the complex ($\sim 30$ to 50 \%) {  . This deviation could also be ascribed to the CR enhancement in the inner Galaxy, that anyway could not exceed 25\% at the galacto-centric distance of the remnant, $\sim$ 6.5 kpc \citep{ringFermi,ringYang}.} Therefore we cannot determine whether the excess is due to the additional component of CRs on top of the CR sea or it is the result of underestimation of the mass of the molecular gas complex.

\begin{table}[]
    \centering
    \begin{tabular}{lll}
    \hline
       Source  &  Parameters & W$_{p} \bigg(\frac{d}{2.2 kpc}\bigg)^{-2}$ [erg]\\
         \hline
         W44 (pp) & $\alpha_1 = 2.40  \pm 0.02 $ &\\
         					&{ $\alpha_2 = 3.87 \pm 0.14$ }& 3.7 $\times 10^{48} \bigg( \frac{n_H}{10~\mathrm{cm}^{-3}}\bigg)^{-1}$\\
               &  $E_{b}= 39\pm$ 3 GeV &\\
         
         \hline
         SE-Source & $ \alpha_1= 1.2\pm 0.2 $ & \\
         & $ \alpha_1= 5.4 \pm 0.8 $ &  0.75 $\times 10^{48} \bigg( \frac{n_H}{7~\mathrm{cm}^{-3}}\bigg)^{-1}$ \\
         &  $ E_{b}= 51\pm 7$ GeV &\\
         \hline
          NW-Source &  $\alpha_1 =2.3 \pm 0.2  $ & \\
          & $\alpha_2=2.64 \pm 0.05  $ &   3.9 $\times 10^{48} \bigg( \frac{n_H}{7~\mathrm{cm}^{-3}}\bigg)^{-1}$\\
          &    $E_{b}= 1.9 \pm 0.8 $ GeV &\\
         \hline
         Surr. Gas  & $ \alpha_1=2.2\pm  0.1$ &\\
         &   $ \alpha_2=2.87\pm  0.08$  &  18 $\times 10^{48} \bigg( \frac{n_H}{2~\mathrm{cm}^{-3}}\bigg)^{-1}$ \\
       	 & $ E_{b}=  20\pm $5  GeV  &  \\
         \hline
         Surr. Gas   				& $ \alpha_1=2.3 \pm 0.4 $ &\\
         (wo Sea) &   $ \alpha_2=2.8 \pm 0.5  $  & 8.5 $\times 10^{48} \bigg( \frac{n_H}{2~\mathrm{cm}^{-3}}\bigg)^{-1}$\\
       			& $ E_{b}= 22\pm 16$  GeV  &  \\
         \hline
    \end{tabular}
    \caption{Parameter derived with  \texttt{naima} fit for the sources of interest. The total proton energy content is calculated above 10 GeV}
    \label{tab:cr_values}
\end{table}{}



%
\subsection{Cosmic Ray content}
We evaluated the content of cosmic ray density from the Cosmic Ray spectra derived with \texttt{naima}, as :
\begin{equation}
  \omega_P = \frac{W_p}{V}= \int E \frac{dN}{dEdV} dE
\end{equation}
both for the two clouds and for the gas complex that surrounds W44. For the two sources, NW-source and SE-source, we find a cosmic ray density above 10 GeV  (corresponding to $\approx$ 1 GeV $\gamma$ rays) of $\omega_{CR}$(NW)= 0.46 $\pm$ 0.14 eV cm$^{-3}$  and  $\omega_{CR}$(SE)= 1.9 $\pm$ 0.6 eV  cm$^{-3}$  respectively, whereas in the surrounding gas:  $\omega_{CR}$(Surr)= 0.31 $\pm$ 0.09 eV cm$^{-3}$. The resulting value are enhanced with respect to the  local value as derived from the AMS02 measurement: $\omega_{CR}$(AMS)= 0.14  eV  cm$^{-3}$, above 10 GeV. 
We see that the CR-density in the clouds is higher than the surrounding medium, sign that they hold a higher concentration of cosmic rays. Remarkably the density SE cloud is enhanced by a factor 6 with respect to the surrounding medium.

\section{Summary}

In this Letter, we present the results of an analysis of almost 10 years of \fermi    observations of the Supernova remnant W44 and its $R \leq 70$~pc ($\leq 2^\circ$) environment.  We confirm the morphology of this middle-age SNR in $\gamma$-rays reported in previous studies. It can be represented as an elliptical ring with inner and outer axes: [0.18$^\circ$,0.3$^\circ$] and [0.13$^\circ$,0.22$^\circ$],respectively. In general, a similar morphology has been reported also by radio observations. 

We derived the SED of W44 over three decades from $\sim$ 100 MeV to $\sim$ 100 GeV (see Fig.\ref{fig:w44SED}). The spectrum is consistent with previous observations \citep{w44fermi,w44agilefermi}. We do not confirm the hardening of the $\gamma$-ray spectrum around 10 GeV as reported by \citep{w44reacc}. Instead,  one can see that the spectral points around 200~GeV lie above the extrapolation of the spectrum from lower energies.  However, an extension of the spectrum at higher energies is fundamental to prove such a tendency above a few hundred GeVs.

Although the systematic uncertainties prevent us from robust conclusions concerning the origin of $\gamma$-ray emission, the very hard $\gamma$-ray spectrum below 1 GeV gives preference to the hadronic origin of radiation.  The alternative interpretation of radiation by electron bremsstrahlung is less likely (see the discussion in Sec. 2). The broadband $\gamma$-ray spectrum over three decades can be explained by hadronic interactions assuming a power-law proton spectrum with an index $\sim 2.4$, and a break (steepening) above few tens of GeV. The break in the proton spectrum suggests that the particles of higher energies have already left the remnant, while GeV particles still reside inside it. 

The location of W44 inside the dense gas environment provides a unique opportunity for the study of the spatial distribution of cosmic rays in this extended region and thus for exploration of the character (geometry) of the escape of accelerated particles from this middle-age SNR. 
Our study shows clear evidence of two extended $\gamma$-ray structures located at two opposite edges of the remnant, along its major axis. The presence of two extended sources in the proximity of W44 has been reported earlier, but the origin of these structures was interpreted as a result of the presence of dense gas clouds \citep{uchiyamaW44}. {  However, 
our study revealed that the emission arising from the entire gas complex, is compatible to the flux
expected from the interactions of the CR Sea, whereas for NW- and SE- sources the content of CRs is 
well above the density of the CR sea}.

If these particles are indeed linked to W44, (one cannot exclude that both objects could appear as background or foreground $\gamma$-ray sources close to W44), then their spatial distribution tells us that the escape of cosmic rays from the shell of W44 occurs anisotropically  {  along the magnetic field lines \citep{w44magn}} . While this scenario needs to be confirmed by further morphological studies and  by comprehensive multi-wavelength modelling,  we notice  that a self-regulated collective escape of particles in the form of  CR \textquotedblleft clouds"  has been predicted by  \cite{Malkov2013}. {  Further investigations are needed to understand the different spectral characteristics of two CR clouds and clarify whether this effect {  i) is due to an intrinsic asymmetry in the shock (e.g. different Mach number throughout the remnant), ii) is the results of propagation effects (as the physical distance of the clouds might be different) or iii) could be caused by irregularities in the interstellar medium (e.g. due to the magnetic field fluctuations \citep{Giacinti}).}}

\appendix 
 \section{MODELING OF THE SOURCES} 

The modeling of the new sources was based on the likelihood test for nested models and on the AIC test for the non-nested models. We define $\mathcal{L}=\log{L}$, where $L$ is the likelihood resulting from the fit. The model with the highest likelihood $L$ is favoured. To evaluate by which extent the model is preferred on another we calculate $\Delta\mathcal{L} = -2(\log{L_0} - \log{L_1})$ : if $\sqrt{\Delta\mathcal{L}}>3$ one model is favoured, if $\sqrt{\Delta\mathcal{L}}<3$ there is not a clearly favoured model. For the non-nested models we define: AIC = $ 2k -2 \mathcal{L}$, where $k$ is the number of parameters of the model; the model with a lower AIC is favoured. Since only relative numbers matter, we calculate a reduced AIC:  AIC*(k*)=2k*-2$\mathcal{L} $ where k* is the difference between the number of parameters in the null hypothesis and the new source model. 

\subsubsection{Modeling of W44}
We firstly studied the morphology and the spectral shape of the supernova remnant, testing different models (see Table \ref{tab:morph_W44}). The favourite shape is an elliptic ring; a model with a shape similar to the one reported in the 4FGL catalog,  shows a higher likelihood, however the difference is not enough to claim a different shape and we preferred to maintain the cataloged shape. For what concerns the spectrum the favorite spectral model results to be a Log-parabola, confirming the model given in the LAT catalog.

\begin{table}[ht!]
    \centering
    \begin{tabular}{lcc}
        \hline
        Morphology ($>$ 1 GeV) & {$\mathcal{L}$} &  \\
            \hline
        Null   hypothesis       & 95 1065 & \\
        Disk           						& 95 6614 &\\
        Ellipse   best   (a=0.25$^{\circ}$ b= 0.375$^{\circ}$)  & 95 4330 &\\
        {\bf Elliptic Ring ( 0.18$^\circ$,0.3$^\circ$ ;0.13$^\circ$ ,0.22$^\circ$ ) }	  						& {\bf 95 7058}&   \\
    	Radio Template (SRT\footnote{\cite{srt}})      			 	  & 95 6770 & \\
       Radio Template (NVSS\footnote{\cite{nvss}})				& 95 6970 &\\ 
        
         \hline \hline
         Spectrum  ($>$ 60 MeV) &{$\mathcal{L}$ } & AIC*     \\
          \hline
          Power Law  & Failed & -- \\
          Smooth Broken Power Law &  987 4900  & -19749788 \\
          Exp Cutoff Power Law &  Failed -- \\
         {\bf Log Parabola} & {\bf 987 5084} & {\bf -19750160 }  \\

         \hline
    \end{tabular}
    \caption{Modeling of the SNR W44. In the upper part the results for the morphological tests, that we conducted starting from 1 GeV, to have a better angular resolution; in the lower part the results for the modeling of the spectrum conducted in the whole energy range ($>$ 60 MeV).}
    \label{tab:morph_W44}
\end{table}{}

\subsubsection{Modeling of the Surroundings}
After fixing the supernova remnant W44, we tested a new morphology for the surrounding sources. We deleted all the sources within 1$^{\circ}$ from the remnant as listed in table \ref{tab:del_sources} and remodeled the emission in that region. 

\begin{table}[ht!]
    \centering
    \begin{tabular}{|ccccc|}
    \hline
         4FGL name & (l,b)$^{\circ}$ & 3FGL name  & Possible Association & New Morphology  \\
         \hline
         \bf{J1852.6+0203} &(34.90,0.67)  & J1852.8+0158 &   YNG PSR candidate\footnote{\cite{Saz}} & Point Source\\
        \bf{J1854.7+0153} & (34.99,0.12) &  -- &  & \multirow{2}{*}{0.4$^{\circ}$-Disk (NW-Source)}\\
       \bf{J1855.8+0150}  & (35.07,-0.14) & -- & -- &  \\
       \bf{J1857.4+0106} & (34.605,0.026) & -- & -- & \multirow{2}{*}{0.15$^{\circ}$-Disk (SE-Source)} \\ 
        \bf{J1857.1+0056} & (34.43,-0.849)& J1857.2+0059 &  YNG PSR candidate\footnote{\cite{Saz}}  & \\ 
       {J1857.6+0143}  & (35.18,-0.59) & J1857.8+0129c  & PSR J1857+0143 & --  \\
       

        \bf{1857.7+0246e} & (36.12,-0.15) &--&  PWN   HESS 1857+026 & 0.3$^{\circ}$ Disk \\
         J1858.3+0209 & (35.63,-0.54) &  \multirow{2}{*}{ J1857.9+0210}& \multirow{2}{*}{MC/SFR \footnote{\cite{Paron,Paredes} } HESS 1857.9+0210} & -- \\
          J1857.6+0212 & (35.60,-0.38) &  & -- &-- \\
       \bf{J1852.4+0037e} & (33.61,0.08) & --& SNR   Kes79 & 0.4$^\circ$-Disk\\
    \hline
    \end{tabular}
    \caption{4FGL sources in the region of W44; the bolded sources are the ones for which we found a new morphology}    \label{tab:del_sources}
\end{table}

In Table \ref{tab:surr_model}, we reported the different configuration that we tried for the emission at the Southern-East and Northern-West edges of the remnant. The best resulting configuration is a 0.15$^\circ$ disk for the Southern emission, and a 0.4$^\circ$-disk and a point-like source for the northern edge. 

\begin{table}[ht!]
    \centering
    \begin{tabular}{llcc}
            \hline
            &  Model & {$\mathcal{L} $} & {AIC*}\\
            
            \hline
            	& Null 					 &  956847  &-- \\ 
 & 2 Pointsources &  957059 &  -1914110\\
    South East              & \bf{1 Disk r=0.15$^\circ$} 				&  \multirow{3}{*}{\bf{957064}}  & \multirow{3}{*}{ {\bf -1914122}}\\
                  &  \ \ \ {\bf TS\_ext =47; } & & \\ 
                    & \ \ \  {\bf l,b = (34.45$^\circ$ , -0.86$^\circ$ ) } & & \\ 
                  \hline
          									  & 1 Disk & 956649 &   -1913292  \\ 
                   								 & 1 PointSource  &  956611 &  -1913218 \\ 
                & Null &   956793 & --\\
                & 2 PointSources &957059 & -1914110 \\
        North West                & {\bf 1 Disk (r=0.4$^{\circ}$)+1 PS} & \multirow{3}{*}{\bf 957081}  &  \multirow{3}{*}{\bf -1914152} \\
                   &  \ \ \ {\bf TS\_ext = 152; } & & \\ 
                    & \ \  \ {\bf l,b = (35.00$^\circ$, 0.04$^\circ$); } & & \\ 
                    
                   & 1 Disk + 1 Disk (TS\_ext $<20$)  &  -- & -- \\
                  
                   \hline

    \end{tabular}
    \caption{The different morphology tested to model the SE and the NW Source. We report the Log-likelihood value for each configuration.}
    \label{tab:surr_model}
\end{table}


\subsection{Modeling of the parent particles with Naima }

We modeled the observed $\gamma$-ray emission with \texttt{naima} version 0.8.1 \citep{naima}. For every source we tested different distribution for the parent particles and determine the best profile using the Bayesian Information Criterion implemented within \texttt{naima}. The model with the lower BIC is favoured.

For W44 we tested 2 different emission mechanism, namely Bremsstrahlung and Pion Decay. In  Tab. \ref{tab:naima_w44} we report all the tested spectra with the relative BIC information. 

\begin{table}[h!]
    \centering
    \begin{tabular}{l|llc}
\hline
    W44           & Model  & Parameters & BIC   \\
                  \hline
                   & Power Law     & $\alpha=2.7355 \pm 0.005 $& 894 \\
    Pion Decay     & Broken Power Law &$\alpha_1=2.40 \pm 0.02$; $\alpha_2= 3.87 \pm 0.14 $;  $E_b= 39 \pm 3$ GeV; & 84 \\
                   & Exponential Cutoff Power Law&$\alpha=2.30 \pm 0.02$; $E_{co}=71 \pm 6$ GeV  &  105 \\
    \hline
                   & Power Law     &  $\alpha=2.36 \pm $ 0.03 & 10308 \\
    Bremsstrahlung & Broken Power Law & $\alpha_1=2.31 \pm 0.03 $ ; $\alpha_2= 3.38\pm 0.06 $; $E_{b}= 6.1 \pm 0.5$ GeV & 153 \\
      ($E^{e}_{min}=600$ MeV)             & Exponential Cutoff Power Law& $\alpha= -0.12 $ ; $E_{co}=1.12 \pm 0.02$ GeV  & 1419  \\
\hline    
    \end{tabular}
    \caption{Model parameters derived with naima for the supernova remnant W44.}
    \label{tab:naima_w44}
\end{table}{}

For the clouds and for the surrounding ISM we assumed the pion decay as dominant emission mechanism. The tested models are reported in Table \ref{tab:naima}

\begin{table}[h!]
    \centering
    \begin{tabular}{l|llc}
\hline
                  & Model  & Parameters & BIC   \\
                  \hline
                    & Power Law                    & $\alpha= 2.36 \pm0.04 $ & 86 \\
    SE Source        & Broken Power Law             & $\alpha_1=1.2 \pm 0.2 ; \alpha_2 =5.4 \pm 0.8$;  $E_b= 51\pm7 $ GeV;  & 22\\
                    & Exponential Cutoff Power Law & $\alpha=1.76 \pm 0.4 $; $E_{co}=74 \pm 35 $ GeV; &  23\\
    \hline
                   & Power Law     & $\alpha=2.61 \pm 0.03 $&  56 \\
    NW Source & Broken Power Law & $\alpha_1=2.3  \pm 0.2 ; \alpha_2 =2.64 \pm 0.05 $; $E_b= 1.9 \pm 0.8 $ & 59 \\
                   & Exponential Cutoff Power Law& $\alpha= 2.61$; $E_{co}=210 \pm 120$ GeV; &  58 \\
      \hline
                   & Power Law     & $\alpha=2.65\pm 0..02 $ & 44 \\
 Surrounding Gas & Broken Power Law & $\alpha_1=2.2 \pm 0.1 $; $\alpha_2= 2.87 \pm 0.08$; $E_{b}=20\pm5$ GeV; & 31\\
                   & Exponential Cutoff Power Law& $\alpha=2.59 \pm 0.04$; $E_{co}=1436 \pm 750$  GeV; & 39 \\

\hline    
    \end{tabular}
    \caption{Model parameters derived with naima for the above cited sources}
    \label{tab:naima}
\end{table}{}

\end{document}